\documentclass[sigconf]{acmart}  

\AtBeginDocument{%
  }

\setcopyright{acmlicensed}
\copyrightyear{2024}
\acmYear{2024}
\acmDOI{XXXXXXX.XXXXXXX}

\acmConference[KDD CUP 2024 OAG-Challenges]{KDD CUP 2024 OAG-Challenges, Paper Source Tracing, Technical Report of Team AoboSama @ KDD CUP 2024. }{August 25--29, 2024}{Barcelona, Spain}

\acmISBN{978-1-4503-XXXX-X/18/06}

\begin{document}

\title{Text-Driven Neural Collaborative Filtering Model for Paper Source Tracing}  

\author{Aobo Xu}
\authornote{Team AoboSama}
\authornote{Leader of Team AoboSama}
\email{upc20xab@s.upc.edu.cn}
\orcid{0000-0003-3392-7374}
\affiliation{%
  \institution{China University of Petroleum}
  \streetaddress{No.66, West Changjiang Road, Huangdao District}
  \city{Qingdao}
  \state{Shandong}
  \country{China}
  \postcode{266580}
}

\author{Bingyu Chang}
\authornotemark[1]
\email{S23080035@s.upc.edu.cn}
\affiliation{%
  \institution{China University of Petroleum}
  \streetaddress{No.66, West Changjiang Road, Huangdao District}
  \city{Qingdao}
  \state{Shandong}
  \country{China}
  \postcode{266580}
}

\author{Qingpeng Liu}
\authornotemark[1]
\email{S23080022@s.upc.edu.cn}
\affiliation{%
  \institution{China University of Petroleum}
  \streetaddress{No.66, West Changjiang Road, Huangdao District}
  \city{Qingdao}
  \state{Shandong}
  \country{China}
  \postcode{266580}
}

\author{Ling Jian}
\authornote{Corresponding author}
\email{bebetter@upc.edu.cn}
\affiliation{%
  \institution{China University of Petroleum}
  \streetaddress{No.66, West Changjiang Road, Huangdao District}
  \city{Qingdao}
  \state{Shandong}
  \country{China}
  \postcode{266580}
}

\renewcommand{\shortauthors}{Aobo Xu et al.}

\begin{abstract}
   Identifying significant references within the complex interrelations of a citation knowledge graph is challenging, which encompasses connections through citations, authorship, keywords, and other relational attributes. The Paper Source Tracing (PST) task seeks to automate the identification of pivotal references for given scholarly articles utilizing advanced data mining techniques. In the KDD CUP OAG-Challenge PST track, we design a recommendation-based framework tailored for the PST task. This framework employs the Neural Collaborative Filtering (NCF) model to generate final predictions. To process the textual attributes of the papers and extract input features for the model, we utilize SciBERT, a pre-trained language model. According to the experimental results, our method achieved a score of 0.37814 on the Mean Average Precision (MAP) metric, outperforming baseline models and ranking 11th among all participating teams. The source code is publicly available at \url{https://github.com/MyLove-XAB/KDDCupFinal}. 
\end{abstract}



\received{20 February 2007}
\received[revised]{12 March 2009}
\received[accepted]{5 June 2009}

\maketitle

\section{Introduction}

\begin{figure}
    \centering
    \includegraphics[width=1\linewidth]{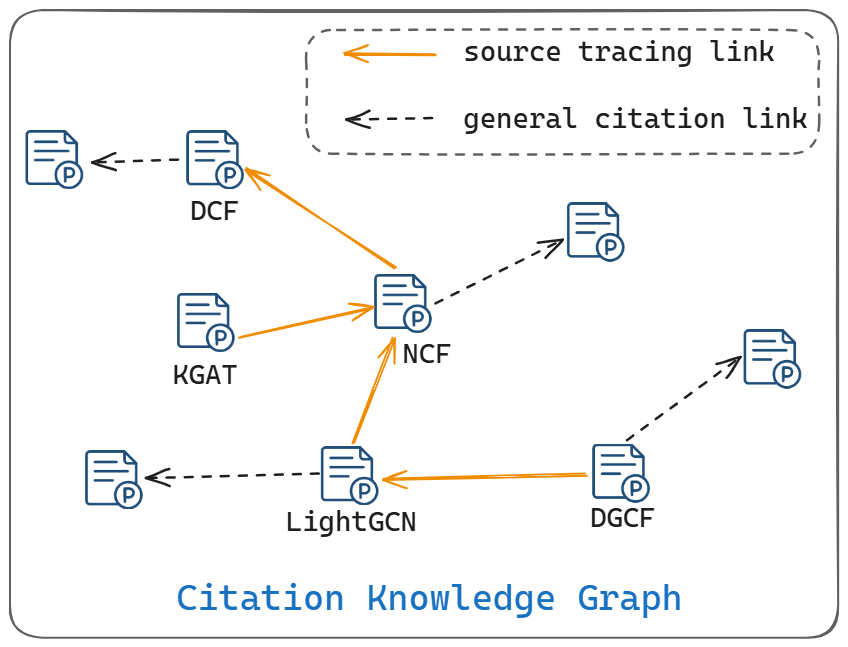}
    \caption{Paper Source Tracing with the Citation Knowledge Graph}
    \label{PST}
\end{figure}

Paper source tracing (PST) is vital for effective knowledge management. By linking common attributes between papers as relational edges, a comprehensive citation knowledge graph is constructed. Successfully addressing the PST task enables the clear visualization of whether paper A primarily inspires or contributes to paper B, as illustrated in Figure~\ref{PST}. Utilizing a citation knowledge graph facilitates the identification of critical references, elucidates the evolution of disciplines and techniques, and foster academic connections and collaborations.
To this end, we propose a model based on Neural Collaborative Filtering (NCF) for the PST task. The model processes the textual attributes of papers obtained from DBLP-citation-network \cite{tang2008arnetminer} (abbreviated as DBLP), with the assistance of SciBERT \cite{beltagy2019scibert}, a pre-trained language model. The proposed model demonstrates suitability for solving the PST task and achieves promising results on the Mean Average Precision (MAP) metric. The source code of our approach is publicly available at \url{https://github.com/MyLove-XAB/KDDCupFinal}. Our contributions are as follows: 
\begin{itemize}
    \item To the best of our knowledge, this is the first recommendation model applied to the paper source tracing task. 
    \item We retrieve the text attributes from knowledge graph and process them with a language model. The experimental results demonstrate promising outcomes. 
\end{itemize}

\section{Related works}
Two primary research fields are pertinent to the PST task from the perspective of the corresponding methodologies: machine learning-based methods and language model-based methods.

\subsection{Machine Learning Based Methods for PST}

The identification of significant references is generally framed as a classification or regression problem. This process involves the manual engineering of features, such as metadata-based parameters \cite{qayyum2019identification}, citation counts\cite{valenzuela2015identifying} \cite{nazir2022important} , author overlap \cite{an2022important}, textual similarity \cite{hassan2017identifying} \cite{nazir2022important}, among others. These features are then utilized for training and inference with various machine learning methods, such as support vector machine, random forest, decision tree, kernel logistic regression, among others \cite{hassan2017identifying} \cite{aljohani2021text} \cite{qayyum2019identification}. Machine learning-based methods offer the advantages of automating the processing of extensive data and efficiently extracting valuable insights for literature analysis. However, these methods may suffer from limitations in accuracy and interpretability, which can be attributed to factors such as data quality and the inherent complexity of the models. 

\subsection{Language Model Based Methods for PST}

Recent advancements in language models have significantly enhanced text processing capabilities. Fine-tuning these models with annotated data has demonstrated promising results for PST task. 
Previous research has explored the integration of text embedding techniques such as Word2Vec \cite{iqbal2023exploiting} and GloVe \cite{safder2023neural}. By framing the PST as a binary classification problem for citation significance, the application of models such as SciBERT, GLM, and Galactica in conjunction with a Multi-Layer Perceptrons (MLPs) classifier has yielded promising results \cite{zhang2024pst}. Building on these advancements, the PST-Bench dataset has been introduced, utilizing pre-trained models like BERT and SciBERT, which further demonstrate the effectiveness of these models in identifying significant references \cite{zhang2024oag}. Language model-based methods are widely used due to their ability in deeply understanding contextual semantics. Nevertheless, the full potential of these models remains largely untapped.

\section{Methodology}

In this section, we first frame the PST task within the context of recommender systems. We then propose a NCF-based model that integrates both the interactions between papers and their references, as well as the textual attributes associated with the papers.

\subsection{Problem Formulation}
\subsubsection{Citation Knowledge Graph}
We define a citation knowledge graph as $\mathcal{G}=(\mathcal{E}, \mathcal{R})$, where $\mathcal{E}$ represents the set of entities, including papers, authors, and other relevant elements, while $\mathcal{R}$ donates the set of relations that describe the interactions between different types of nodes. Each paper node is associated with various text attributes, such as the paper title, abstract and body. 

\subsubsection{Recommendation Modeling}
A conventional approach to address the PST task involves modeling it as a matching problem, where similarity between the query paper and potential references is calculated. Similarly, recommender systems operate by filtering items based on user preferences through similarity measures. Given the conceptual overlap between the PST task and recommendation systems, we frame the PST task as a recommendation problem. From the perspective of recommender systems, the query papers can be conceptualized as "users" and references as "items". In this framework, "citation" relations represent interactions between "users" and "items", with "critical citations" interpreted as positive interactions. By reformulating the problem as a recommendation task, we can leverage the family of algorithms used in recommender systems. Specifically, we focus on predicting the probability of a "critical citation" for each paper-reference pair $(p, r)$, $p, r \in \mathcal{E}$. To this end, we employ collaborative filtering algorithms, which, despite their simplicity, have proven to be effective. 

\subsection{NCF-based PST Model with SciBERT}
\begin{figure*}[t]
    \centering
    \includegraphics[width=1 \textwidth]{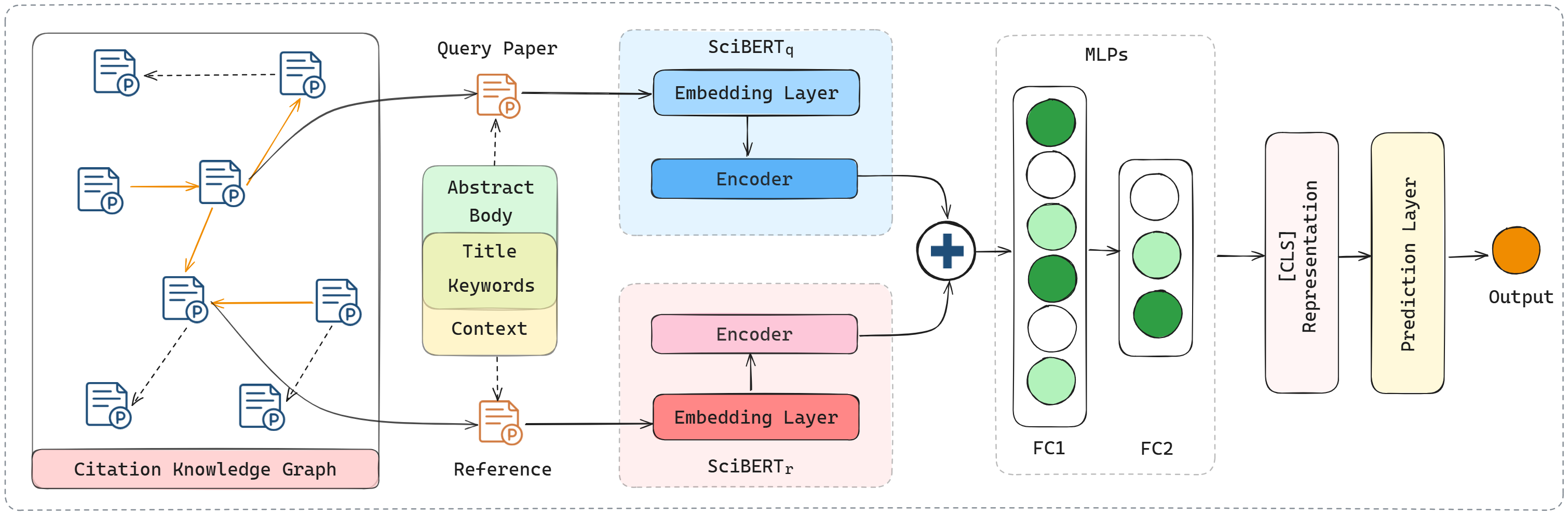}
    \caption{NCF-SciBERT Model for PST Task}
    \label{model}
\end{figure*}

We employ the Neural Collaborative Filtering (NCF) model, which processes the features of papers and references separately in two channels, subsequently computing their mutual similarity as the prediction output. Drawing inspiration from the potential of language models in text-attributed graphs \cite{chen2024exploring}, we utilize a pre-trained language model, specifically SciBERT, to process the textual attributes of the papers and references. The SciBERT module comprises 12 BERT layers, each implementing attention and feed-forward network (FFN) mechanisms. Two separate SciBERT modules encode the query paper and reference inputs independently, after which the [CLS] token representations from the outputs are merged and fed into Multi-Layer Perceptrons (MLPs) to calculate the prediction values. The model architecture is depicted as Figure~\ref{model}. Denoting the inputs of query paper and reference as $q_{in}$ and $r_{in}$ respectively, the complete process of the model is as follows:
\begin{equation}
  q_h = \mathrm{SciBERT_q}(q_{in}),
  \label{eq1}
\end{equation}
\begin{equation}
  r_h = \mathrm{SciBERT_r}(r_{in}),
  \label{eq2}
\end{equation}
\begin{equation}
  h = \mathrm{concat}(q_h[CLS], r_h[CLS]),
  \label{eq3}
\end{equation}
\begin{equation}
  x = \mathrm{MLPs}(h),
  \label{eq4}
\end{equation}
\begin{equation}
  o = \mathrm{softmax}(\mathrm{F}(x)),
  \label{eq5}
\end{equation}
where $\mathrm{F}$ represents the final prediction layer and the $\mathrm{softmax}$ operation converts the output into probability predictions. By training a sophisticated recommender model, the model is competent for solving the PST task. 

\section{Experiments}

In this section, we perform comparison experiments to assess the suitability of the NCF-SciBERT model for the PST tasks.

\subsection{Dataset}
We conduct experiments on the PST dataset provided by the organizer. The text input for the query papers includes the title, abstract, keywords, and the first 500 characters of the body. For the references, the inputs comprise the title, keywords, and the contextual text in which each reference appears within the full text. Negative sampling is employed with a positive-to-negative sample ratio of 1:10. From the total of 788 training papers, we construct 8905 samples. The dataset is divided into training and validation subsets with a ratio of 8:2.

\subsection{Experimental Results}
The model is trained on a NVIDIA A100-PCIE-40GB GPU. The total number of model parameters is approximately $2.3 \times 10 ^ 9$. For hyperparameters, the batch size is set to 16, and the maximum length for both paper and reference inputs is 512 tokens. The model's memory usage does not exceed 36 GB. We employ the ADAM optimizer with an initial learning rate of \(1 \times 10^{-5}\).

We evaluate the final performance of our model using the test data provided by the organizer. As baseline comparisons, we select the Random Forest, ProNE, and SciBERT methods proposed in \cite{zhang2024pst}. Table~\ref{tab1} presents the results of various methods evaluated using the Mean Average Precision (MAP) metric. Our proposed model (NCF-SciBERT) achieves 0.37814 in MAP score, representing a significant improvement over the baselines, with a 28.23\% enhancement relative to the best-performing baseline. 

To demonstrate the significance of textual attributes in the task, we design an ablation experiment by introducing an NCF-id method, which excludes textual attributes and uses only the paper ID as input. The ablation results shows the text attribute greatly improve the performance, as shown in Table~\ref{tab1}. Furthermore, the improvement of the NCF-SciBERT model over the SciBERT model alone suggests that the NCF architecture has a positive impact on the PST task.

\begin{table}
  \caption{Results of Different Methods on MAP Metric}
  \label{tab1}
  \begin{tabular}{cccc}
  \toprule 
    Method & Valid MAP & Test MAP & Description  \\
    \hline
    Random Forest & 0.21551 & 0.18020 & Manual-feature based \\
    ProNE Baseline &  0.19104 & 0.16387 & GNN based\\ 
    SciBERT & 0.29489 & 0.22237 & LM based\\ 
    NCF-id & 0.18111 & 0.15569 & No text attribute\\ 
    NCF-SciBERT & \textbf{0.39065} &  \textbf{0.37814} & Our method\\
  \bottomrule  
\end{tabular}
\end{table}

\section{Conclusion}

In this paper, we formalize the paper source tracing task as a problem within the domain of recommender systems. We integrate the Neural Collaborative Filtering model with a language model, specifically SciBERT, by leveraging textual attributes retrieved from the citation knowledge graph. The experimental results demonstrate that the model achieves a MAP score of 0.37814 on the test dataset, which is a notable improvement over baseline methods and ranking 11th among all participating teams. Future work will explore the application of graph reasoning techniques to further address this task.

\begin{acks}
This work was supported by the National Key Research and Development Program of China under Grant Nos. 2021YFA1000100 and 2021YFA1000104, Laboratory Project of Higher Education Institutions in Shandong Province-Energy System Intelligent Management and Policy Simulation Laboratory at China University of Petroleum, and Youth Innovation Team of Higher Education Institutions in Shandong Province-Data Intelligence Innovation Team at China University of Petroleum.
\end{acks}

\bibliographystyle{ACM-Reference-Format}
\bibliography{sample-base}


\begin{thebibliography}{13}


\ifx \showCODEN    \undefined \def \showCODEN     #1{\unskip}     \fi
\ifx \showDOI      \undefined \def \showDOI       #1{#1}\fi
\ifx \showISBNx    \undefined \def \showISBNx     #1{\unskip}     \fi
\ifx \showISBNxiii \undefined \def \showISBNxiii  #1{\unskip}     \fi
\ifx \showISSN     \undefined \def \showISSN      #1{\unskip}     \fi
\ifx \showLCCN     \undefined \def \showLCCN      #1{\unskip}     \fi
\ifx \shownote     \undefined \def \shownote      #1{#1}          \fi
\ifx \showarticletitle \undefined \def \showarticletitle #1{#1}   \fi
\ifx \showURL      \undefined \def \showURL       {\relax}        \fi
\providecommand\bibfield[2]{#2}
\providecommand\bibinfo[2]{#2}
\providecommand\natexlab[1]{#1}
\providecommand\showeprint[2][]{arXiv:#2}

\bibitem[Aljohani et~al\mbox{.}(2021)]%
        {aljohani2021text}
\bibfield{author}{\bibinfo{person}{Naif~Radi Aljohani}, \bibinfo{person}{Ayman Fayoumi}, {and} \bibinfo{person}{Saeed-Ul Hassan}.} \bibinfo{year}{2021}\natexlab{}.
\newblock \showarticletitle{An in-text citation classification predictive model for a scholarly search system}.
\newblock \bibinfo{journal}{\emph{Scientometrics}} \bibinfo{volume}{126}, \bibinfo{number}{7} (\bibinfo{year}{2021}), \bibinfo{pages}{5509--5529}.
\newblock


\bibitem[An et~al\mbox{.}(2022)]%
        {an2022important}
\bibfield{author}{\bibinfo{person}{Xin An}, \bibinfo{person}{Xin Sun}, {and} \bibinfo{person}{Shuo Xu}.} \bibinfo{year}{2022}\natexlab{}.
\newblock \showarticletitle{Important citations identification with semi-supervised classification model}.
\newblock \bibinfo{journal}{\emph{Scientometrics}} \bibinfo{volume}{127}, \bibinfo{number}{11} (\bibinfo{year}{2022}), \bibinfo{pages}{6533--6555}.
\newblock


\bibitem[Beltagy et~al\mbox{.}(2019)]%
        {beltagy2019scibert}
\bibfield{author}{\bibinfo{person}{Iz Beltagy}, \bibinfo{person}{Kyle Lo}, {and} \bibinfo{person}{Arman Cohan}.} \bibinfo{year}{2019}\natexlab{}.
\newblock \showarticletitle{SciBERT: A Pretrained Language Model for Scientific Text}. In \bibinfo{booktitle}{\emph{Proceedings of the 2019 Conference on Empirical Methods in Natural Language Processing and the 9th International Joint Conference on Natural Language Processing (EMNLP-IJCNLP)}}. \bibinfo{pages}{3615--3620}.
\newblock


\bibitem[Chen et~al\mbox{.}(2024)]%
        {chen2024exploring}
\bibfield{author}{\bibinfo{person}{Zhikai Chen}, \bibinfo{person}{Haitao Mao}, \bibinfo{person}{Hang Li}, \bibinfo{person}{Wei Jin}, \bibinfo{person}{Hongzhi Wen}, \bibinfo{person}{Xiaochi Wei}, \bibinfo{person}{Shuaiqiang Wang}, \bibinfo{person}{Dawei Yin}, \bibinfo{person}{Wenqi Fan}, \bibinfo{person}{Hui Liu}, {et~al\mbox{.}}} \bibinfo{year}{2024}\natexlab{}.
\newblock \showarticletitle{Exploring the potential of large language models (llms) in learning on graphs}.
\newblock \bibinfo{journal}{\emph{ACM SIGKDD Explorations Newsletter}} \bibinfo{volume}{25}, \bibinfo{number}{2} (\bibinfo{year}{2024}), \bibinfo{pages}{42--61}.
\newblock


\bibitem[Hassan et~al\mbox{.}(2017)]%
        {hassan2017identifying}
\bibfield{author}{\bibinfo{person}{Saeed-Ul Hassan}, \bibinfo{person}{Anam Akram}, {and} \bibinfo{person}{Peter Haddawy}.} \bibinfo{year}{2017}\natexlab{}.
\newblock \showarticletitle{Identifying important citations using contextual information from full text}. In \bibinfo{booktitle}{\emph{2017 ACM/IEEE joint conference on digital libraries (JCDL)}}. IEEE, \bibinfo{pages}{1--8}.
\newblock


\bibitem[Iqbal et~al\mbox{.}(2023)]%
        {iqbal2023exploiting}
\bibfield{author}{\bibinfo{person}{Arshad Iqbal}, \bibinfo{person}{Abdul Shahid}, \bibinfo{person}{Muhammad Roman}, \bibinfo{person}{Tanveer Afzal}, {and} \bibinfo{person}{Muhammad Yahya}.} \bibinfo{year}{2023}\natexlab{}.
\newblock \showarticletitle{Exploiting Contextual Word Embedding for Identification of Important Citations: Incorporating Section-Wise Citation Counts and Metadata Features}.
\newblock \bibinfo{journal}{\emph{IEEE Access}} (\bibinfo{year}{2023}).
\newblock


\bibitem[Nazir et~al\mbox{.}(2022)]%
        {nazir2022important}
\bibfield{author}{\bibinfo{person}{Shahzad Nazir}, \bibinfo{person}{Muhammad Asif}, \bibinfo{person}{Shahbaz Ahmad}, \bibinfo{person}{Hanan Aljuaid}, \bibinfo{person}{Rimsha Iftikhar}, \bibinfo{person}{Zubair Nawaz}, {and} \bibinfo{person}{Yazeed~Yasin Ghadi}.} \bibinfo{year}{2022}\natexlab{}.
\newblock \showarticletitle{Important citation identification by exploding the sentiment analysis and section-wise in-text citation weights}.
\newblock \bibinfo{journal}{\emph{IEEE Access}}  \bibinfo{volume}{10} (\bibinfo{year}{2022}), \bibinfo{pages}{87990--88000}.
\newblock


\bibitem[Qayyum and Afzal(2019)]%
        {qayyum2019identification}
\bibfield{author}{\bibinfo{person}{Faiza Qayyum} {and} \bibinfo{person}{Muhammad~Tanvir Afzal}.} \bibinfo{year}{2019}\natexlab{}.
\newblock \showarticletitle{Identification of important citations by exploiting research articles’ metadata and cue-terms from content}.
\newblock \bibinfo{journal}{\emph{Scientometrics}}  \bibinfo{volume}{118} (\bibinfo{year}{2019}), \bibinfo{pages}{21--43}.
\newblock


\bibitem[Safder et~al\mbox{.}(2023)]%
        {safder2023neural}
\bibfield{author}{\bibinfo{person}{Iqra Safder}, \bibinfo{person}{Momin Ali}, \bibinfo{person}{Naif~Radi Aljohani}, \bibinfo{person}{Raheel Nawaz}, {and} \bibinfo{person}{Saeed-Ul Hassan}.} \bibinfo{year}{2023}\natexlab{}.
\newblock \showarticletitle{Neural machine translation for in-text citation classification}.
\newblock \bibinfo{journal}{\emph{Journal of the Association for Information Science and Technology}} \bibinfo{volume}{74}, \bibinfo{number}{10} (\bibinfo{year}{2023}), \bibinfo{pages}{1229--1240}.
\newblock


\bibitem[Tang et~al\mbox{.}(2008)]%
        {tang2008arnetminer}
\bibfield{author}{\bibinfo{person}{Jie Tang}, \bibinfo{person}{Jing Zhang}, \bibinfo{person}{Limin Yao}, \bibinfo{person}{Juanzi Li}, \bibinfo{person}{Li Zhang}, {and} \bibinfo{person}{Zhong Su}.} \bibinfo{year}{2008}\natexlab{}.
\newblock \showarticletitle{Arnetminer: extraction and mining of academic social networks}. In \bibinfo{booktitle}{\emph{Proceedings of the 14th ACM SIGKDD international conference on Knowledge discovery and data mining}}. \bibinfo{pages}{990--998}.
\newblock


\bibitem[Valenzuela et~al\mbox{.}(2015)]%
        {valenzuela2015identifying}
\bibfield{author}{\bibinfo{person}{Marco Valenzuela}, \bibinfo{person}{Vu Ha}, {and} \bibinfo{person}{Oren Etzioni}.} \bibinfo{year}{2015}\natexlab{}.
\newblock \showarticletitle{Identifying meaningful citations}. In \bibinfo{booktitle}{\emph{Workshops at the twenty-ninth AAAI conference on artificial intelligence}}.
\newblock


\bibitem[Zhang et~al\mbox{.}(2024a)]%
        {zhang2024pst}
\bibfield{author}{\bibinfo{person}{Fanjin Zhang}, \bibinfo{person}{Kun Cao}, \bibinfo{person}{Yukuo Cen}, \bibinfo{person}{Jifan Yu}, \bibinfo{person}{Da Yin}, {and} \bibinfo{person}{Jie Tang}.} \bibinfo{year}{2024}\natexlab{a}.
\newblock \showarticletitle{PST-Bench: Tracing and Benchmarking the Source of Publications}.
\newblock \bibinfo{journal}{\emph{arXiv preprint arXiv:2402.16009}} (\bibinfo{year}{2024}).
\newblock


\bibitem[Zhang et~al\mbox{.}(2024b)]%
        {zhang2024oag}
\bibfield{author}{\bibinfo{person}{Fanjin Zhang}, \bibinfo{person}{Shijie Shi}, \bibinfo{person}{Yifan Zhu}, \bibinfo{person}{Bo Chen}, \bibinfo{person}{Yukuo Cen}, \bibinfo{person}{Jifan Yu}, \bibinfo{person}{Yelin Chen}, \bibinfo{person}{Lulu Wang}, \bibinfo{person}{Qingfei Zhao}, \bibinfo{person}{Yuqing Cheng}, \bibinfo{person}{Tianyi Han}, \bibinfo{person}{Yuwei An}, \bibinfo{person}{Dan Zhang}, \bibinfo{person}{Weng~Lam Tam}, \bibinfo{person}{Kun Cao}, \bibinfo{person}{Yunhe Pang}, \bibinfo{person}{Xinyu Guan}, \bibinfo{person}{Huihui Yuan}, \bibinfo{person}{Jian Song}, \bibinfo{person}{Xiaoyan Li}, \bibinfo{person}{Yuxiao Dong}, {and} \bibinfo{person}{Jie Tang}.} \bibinfo{year}{2024}\natexlab{b}.
\newblock \showarticletitle{OAG-Bench: A Human-Curated Benchmark for Academic Graph Mining}. In \bibinfo{booktitle}{\emph{Proceedings of the 30th ACM SIGKDD Conference on Knowledge Discovery and Data Mining}}.
\newblock


\end{thebibliography}





\end{document}